\documentclass[aps, twocolumn, groupedaddress,superscriptaddress, showpacs,pre,nofootinbib]{revtex4-1}
\usepackage{graphicx}
\usepackage[title]{appendix}
\usepackage{array}
\usepackage{setspace,transparent}
\usepackage{color}

\usepackage{fontenc,subcaption,ragged2e,amsmath}
\usepackage{mathtools,amssymb,amssymb,nicefrac,tikz}
\usepackage{soul}
\usepackage[font=small,labelfont=bf]{caption}
\usepackage[utf8]{inputenc}

\DeclareCaptionJustification{justified}{\justifying}
\newcommand{\RN}[1]{%
  \textup{\uppercase\expandafter{\romannumeral#1}}%
}
\DeclareMathAlphabet{\mathpzc}{OT1}{pzc}{m}{it}
\DeclareCaptionJustification{justified}{\justifying}
\begin{document}

\title{Fractal fluctuations at mixed-order transitions in interdependent networks}
\author{Bnaya Gross}
\thanks{Correspondence should be addressed to:\\ bnaya.gross@gmail.com, bonamassai@ceu.edu \\ B.G. \& I.B. contributed equally to this work.}
\affiliation{Department of Physics, Bar Ilan University, Ramat Gan, Israel}
\author{Ivan Bonamassa}
\affiliation{Department of Physics, Bar Ilan University, Ramat Gan, Israel}
\affiliation{Department of Network and Data Science, CEU, Quellenstrasse 51, A-1100 Vienna, Austria}
\author{Shlomo Havlin}
\affiliation{Department of Physics, Bar Ilan University, Ramat Gan, Israel}
\date{\today}

\begin{abstract}
    We study the geometrical features of the order parameter's fluctuations near the critical point of mixed-order phase transitions in randomly interdependent spatial networks. In contrast to continuous transitions, where the structure of the order parameter at criticality is fractal, in mixed-order transitions the structure of the order parameter is known to be compact. Remarkably, we find that although being compact, the fluctuations of the order parameter close to mixed-order transitions are fractal up to a well-defined correlation length $\xi'$, which diverges when approaching the critical threshold. We characterize the self-similar nature of these critical fluctuations through their fractal dimension, $d_f'=3d/4$, and correlation length exponent, $\nu'=2/d$, where $d$ is the dimension of the system. By means of percolation and magnetization, we demonstrate that $d_f'$ and $\nu'$ are independent on the symmetry of the underlying process for any $d$ of the underlying networks. 
\end{abstract}

\maketitle

\emph{\underline{Introduction}}-- Critical phenomena are fundamental features of phase transitions, showing universal behaviours that emerge in the vicinity of the critical point~\cite{domb2000phase,stanley1971phase}. 
In second-order transitions, these phenomena are typically reflected in the scaling relations between critical exponents~\cite{bunde1991fractals,stauffer_book} as well as in the fractal geometry of the order parameter's structure at criticality~\cite{stauffer1979scaling,coniglio1982cluster,coniglio1989fractal}. 
When first-order transitions are considered, these universal features become more cumbersome: the order parameter's structure is typically compact at the transition threshold and the critical singularity remains hidden behind metastable fluctuations, hiding hence its critical properties~\cite{binder1987theory}. 
Hybrid or mixed-order transitions~\cite{parisi2008k, bar2014mixed, boccaletti2016explosive, alert2017mixed}, i.e.\ transitions whose order parameter has both a discontinuity (hence, it is compact~\cite{klein1981percolation}) and a square-root singularity at the critical point, have received much attention in recent years due to their mixed nature, raising the perspective of unveiling novel universal properties. 
\par
In this regard, interdependent networks stem as a suitable venue~\cite{buldyrev2010catastrophic,gao2012networks} for the theoretical understanding of mixed-order transitions, while concurrently emphasizing their importance for a broad and interdisciplinary range of real-world systems~\cite{rosato-criticalinf2008, klosik2017interdependent, rocha2018cascading}. 
Interdependent systems, in fact, typically undergo hybrid structural/functional transitions due to cascading failures~\cite{danziger2019dynamic}, whose properties generally depend on the topological features of the interacting networks~\cite{bashan2013extreme, nicosia2013growing} as well as on the range and fraction of dependency links~\cite{parshani2010interdependent, danziger2016effect}. 
Randomly interdependent networks, in particular, host mixed-order percolation transitions when coupling random or $2$-dimensional networks~\cite{li2012cascading, bashan2013extreme}, emerging then as tailored models to study their critical properties at any dimension. 

\begin{figure}[b]
	\centering
	\includegraphics[width=0.9\linewidth]{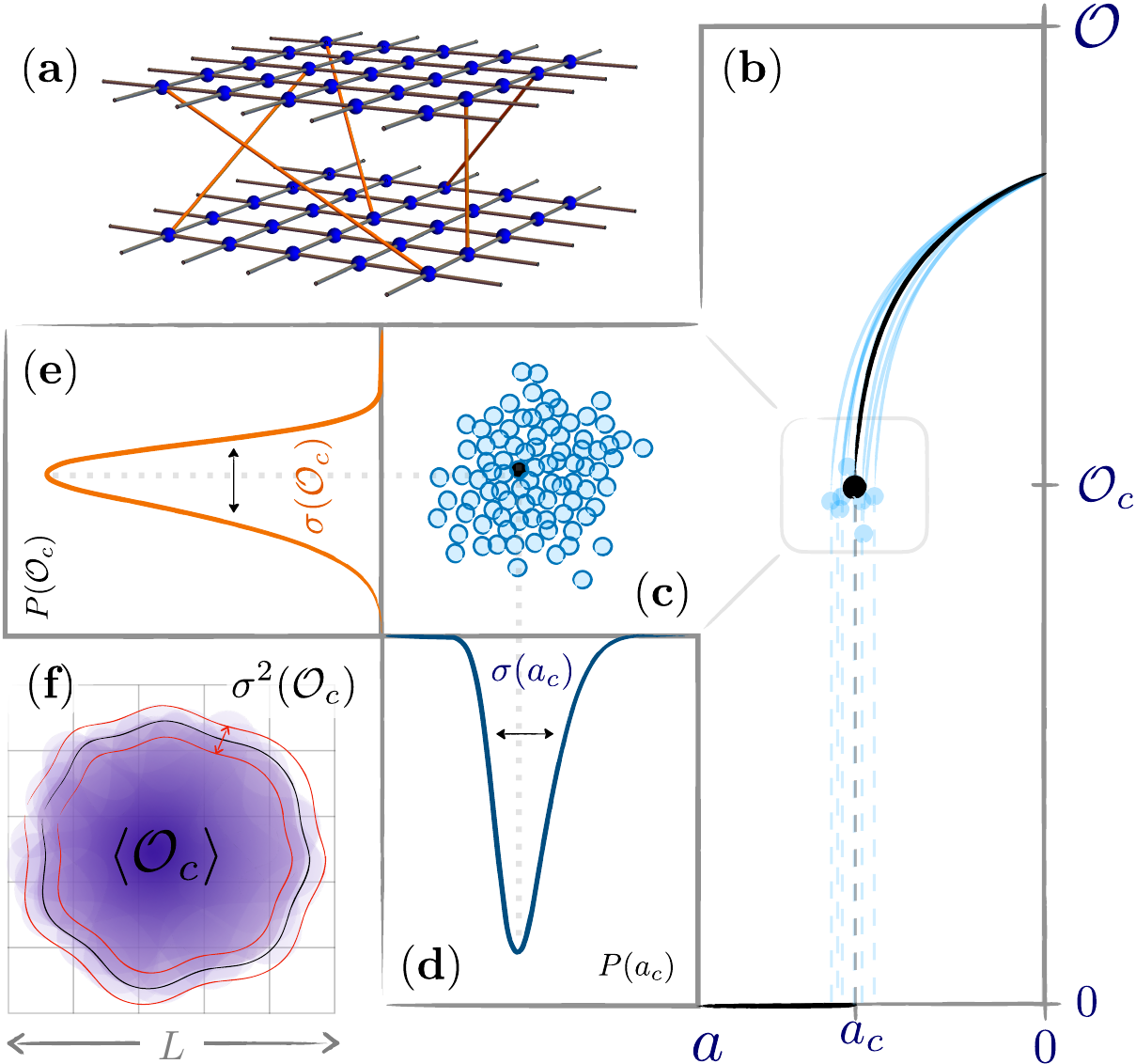}
	\caption{\small \textbf{Fluctuations at a hybrid transition} ( Color online) 
	(\textbf{a}) Illustration of the model of randomly interdependent $d$-dimensional networks (here $d=2$) studied in this Letter, featuring short-range connectivity links (gray links) and long-range dependency links (orange couplings). 
	\textbf{(b)} Each realization of a hybrid transition (see (\textbf{c}) for a zoom-in the bounded region) has its own critical threshold, $a_c$, and critical mass, $\mathcal{O}_c$, whose distributions follow a certain profile, as shown in (\textbf{d}) for $P(a_c)$ and in (\textbf{e}) for $P(\mathcal{O}_c)$.
	(\textbf{f}) Illustration of the fluctuating values of the critical mass, their mean value, $\langle\mathcal{O}_c\rangle$, and their statistical variation $\sigma^2(\mathcal{O}_c)$. 
	} 
	\label{fig:Illustration}	
\end{figure}

\begin{figure*}
	\centering
	\includegraphics[width=0.7\linewidth]{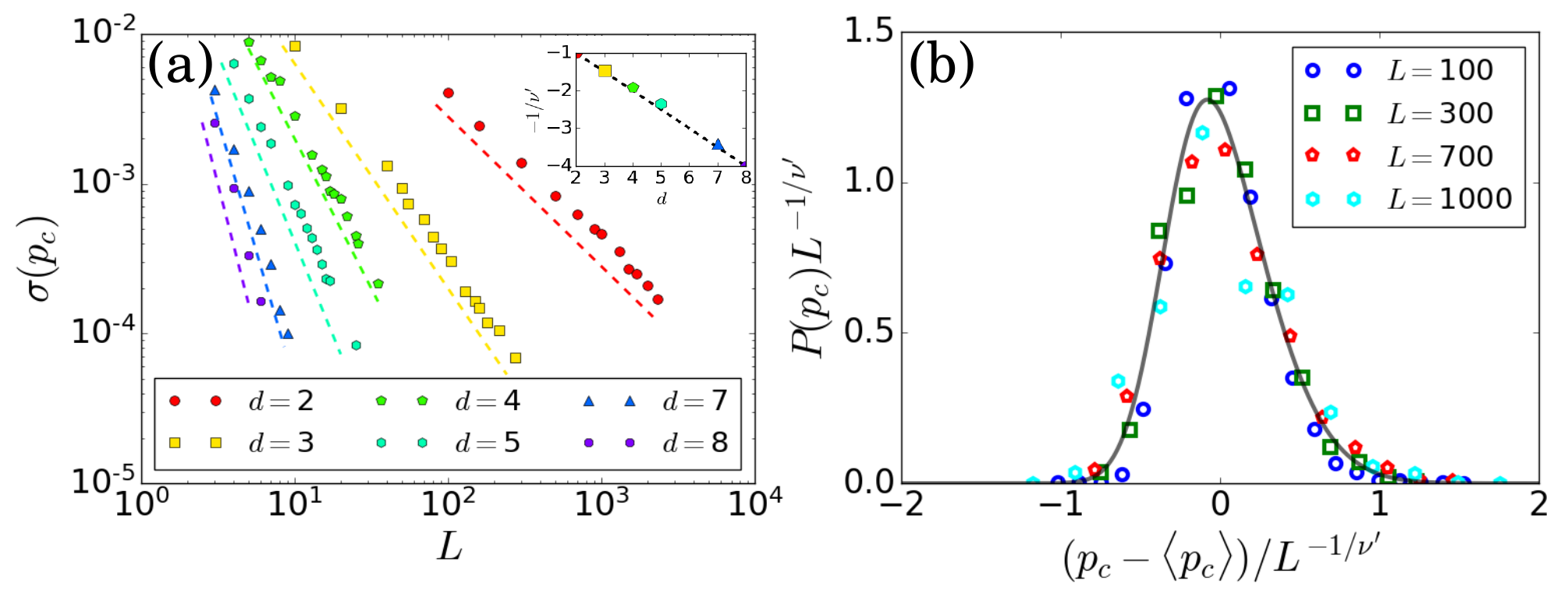}	
	\caption{\textbf{Fluctuations correlation length} (Color online) 
	(\textbf{a}) Simulations of the scaling of $\sigma(p_c)$ with $L$ have, for all studied dimensions ($d=2-8$). We find excellent agreement between simulations and the scaling relation in Eq.~\eqref{eq: nuprime}.
	(Inset) The dependence of $\nu'$ on the dimension $d$ of the underlying lattices follows the relation $\nu' = 2/d$. 
	(\textbf{b}) The distribution of $p_c$ (here, $d=2$) fits a skewed Gaussian and follow the scaling in Eq.~\eqref{eq: pc_dist} (black line) with $\gamma_1 \simeq 0.465$ and $\kappa \simeq 0.315$.}
	\label{fig:nuprime}	
\end{figure*}

\par
In this Letter, we study the critical fluctuations of the order parameter, $\mathcal{O}$, at the mixed-order transition of randomly interdependent networks.  
We find that, although the structure of $\mathcal{O}$ is always compact, its \textit{fluctuations} are {\em self-similar} and are characterized by a well-defined fractal dimension $d_f'$ until a correlation length, $\xi'$, which diverges at the hybrid transition threshold, $a_c$, as 
\begin{equation}
    \xi' \sim |a - a_c|^{-\nu'},\qquad \nu'=2/d,
    \label{eq: xi_prime}
\end{equation}
where $a$ is a control parameter.  
We demonstrate this in both interdependent percolation and interdependent magnetization processes in $d$-dimensional lattices, where we show that: {\em 1}) the exponents $\nu'$ and $d_f'$ are independent on the underlying process, {\em 2}) their values are valid for any dimension $d\geq2$ (i.e.\ the upper and lower critical dimension for hybrid transition is $d_c=2$), and {\em 3}) they satisfy the hyperscaling relation~\cite{ma2018modern}. 
Building on the above, we put forward the hypothesis that fractal fluctuations are a universal property of hybrid phase transitions and we support this claim by developing and testing a unifying scaling theory for the order parameter's fluctuations in the vicinity of the critical point. 
\par

\underline{\textit{Model and main results}}-- Our model consists of two randomly interdependent $d$-dimensional lattices of size $N = L^d$. We construct the dependency links between the layers by randomly pairing the functional states of the two lattices' sites. In such way, we generate a multilayer system with short-range connectivity and long-range dependency (Fig.~\ref{fig:Illustration}\textbf{a}). We consider the two networks to be fully interdependent i.e.\ each node in one layer depends on the state (in what follows, percolation and magnetization) of a single node in the other layer~\cite{li2012cascading}.
\par
By collecting a large sample~\cite{note1} of independent realizations of the hybrid phase transitions reported in both models (Fig.~\ref{fig:Illustration}\textbf{b},\textbf{c}), we study the fluctuations of their critical thresholds, $\sigma^2(a_c) = \langle a_c^2 \rangle - \langle a_c \rangle^2$, and of their order parameter's critical mass, $\sigma^2(\mathcal{O}_c) = \langle \mathcal{O}_c^2 \rangle - \langle \mathcal{O}_c \rangle^2$. 
Both quantities are obtained respectively from the distributions $P(a_c)$ (Fig.~\ref{fig:Illustration}\textbf{d}) and $P(\mathcal{O}_c)$ (Fig.~\ref{fig:Illustration}\textbf{e}). Following a method introduced by Levishtein {\em et al.}~\cite{levinshtein1975english} and later discussed by Stauffer~\cite{stauffer_book} (see also \cite{bunde1991fractals}), we determine the correlation length critical exponent introduced in Eq.~\eqref{eq: xi_prime} by finite-size scaling as 
\begin{equation}\label{eq: nuprime}
    \sigma(a_c)\sim L^{-1/\nu'},\qquad \nu'=2/d.  
\end{equation} 
\par 
A fundamental question then arises: What is the physical role played by a diverging correlation length, Eq.~\eqref{eq: xi_prime}, at a mixed-order phase transition? 
\par
\begin{figure*}
	\centering
	\includegraphics[width=\linewidth]{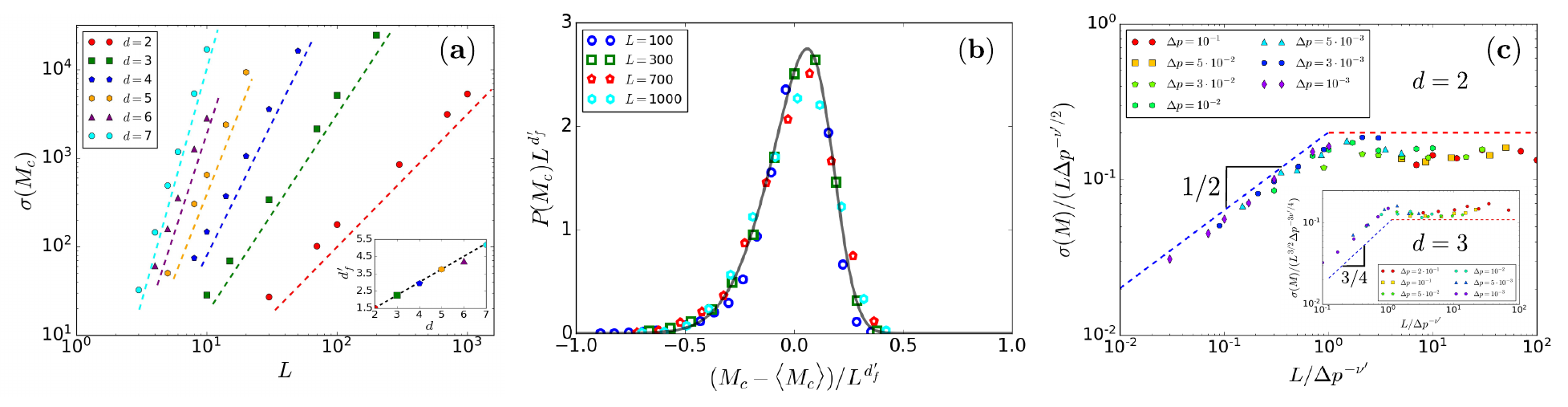}
	\caption{\textbf{Fractal fluctuations} (Color online) 
	(\textbf{a}) Simulations of the scaling of $\sigma(M_c)$ with $L$ show excellent agreement\vspace*{-0.075cm} with Eq.~\eqref{eq: fractal_dimension} where $d_f' = 3d/4$ (see inset). 
	\textbf{(b)} Data collapse of $P(M_c)$ at $d=2$ under the scaling $P(M_c) L^{d_f'} \sim \mathcal{F}[(M_c-\langle M_c \rangle)/L^{d_f'}]$ (black line), where $\mathcal{F}$ is a skewed Gaussian as in Eq.~\eqref{eq: pc_dist}, now with $\gamma_1 \simeq -0.607$ and $\kappa \simeq 0.449$. 
	(\textbf{c}) Close to the hybrid percolation threshold, the MGCC's fluctuations are fractal up to scales below $\xi'$ (Eq.~\eqref{eq: xi_prime}) and non-fractal otherwise, as described by the universal scaling function in Eq.~\eqref{eq :scaling}. Results are shown for $d = 2$ and for $d = 3$ in the inset.}
	\label{fig:dfprime}	
\end{figure*}
\begin{figure*}
	\centering
	\includegraphics[width=0.7\linewidth]{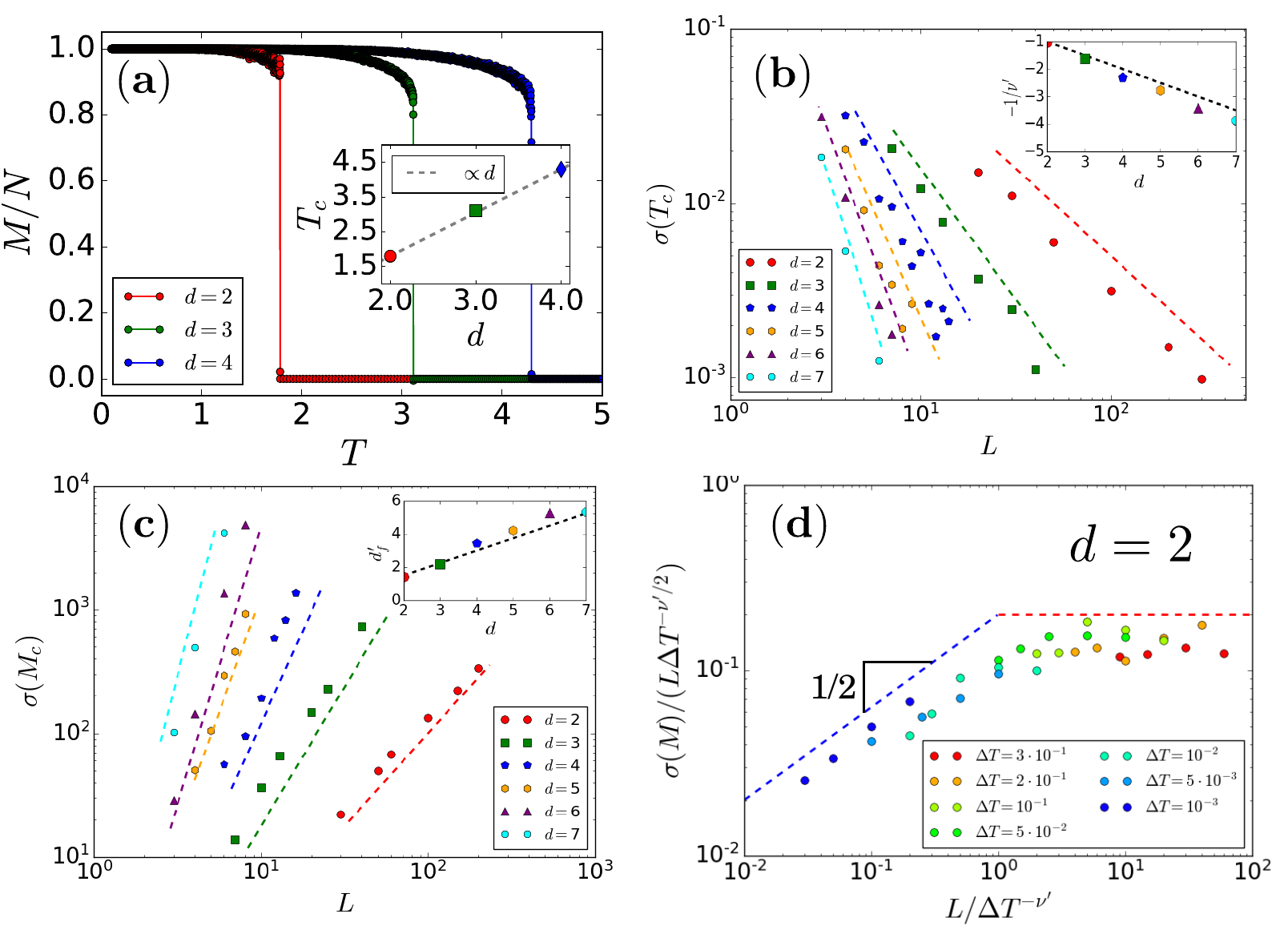}\vspace*{-0.3cm}
	\caption{\textbf{Fractal fluctuations in interdependent magnetization} (Color online) 
	(\textbf{a}) Mixed-order transition in randomly interdependent Ising $d$-dimensional lattices (here $d=2,3,4$) measured via the magnetic density, $M/N$, as function of temperature, $T$. 
	(Inset) The critical temperature $T_c$ scales linearly with the lattices' dimension $d$. 
	(\textbf{b}) Simulations of the scaling of $\sigma(T_c)$ with $L$ show excellent agreement with Eq.~\eqref{eq: nuprime} where $\nu' = 2/d$ as shown in the inset. 
	(\textbf{c}) At criticality $\xi'$ diverges and the fluctuations of the MGCC are fractal in all length scales and follow Eq.~\eqref{eq: fractal_dimension} with $d_f' = 3d/4$ as shown in the inset. 
	(\textbf{d}) Fluctuations are fractal up to $\xi'$ (Eq.~\eqref{eq: xi_prime}) and non-fractal above it, confirming the scaling in Eq.~\eqref{eq :scaling}, here shown with $d = 2$.}
	\label{fig:spins}	
\end{figure*}

In analogy with continuous transitions, where the order parameter (near criticality) is fractal below the correlation length \cite{bonamassa2019criticalstretch}, we show in what follows that at mixed-order transitions the critical {\em fluctuations} of the order parameter's mass (Fig.~\ref{fig:Illustration}\textbf{f}) themselves are self-similar up to $\xi'$, with a well-defined fractal dimension, $d_f'$, given by
\begin{equation}\label{eq: fractal_dimension}
    \sigma(\mathcal{O}_c)\sim L^{d_f'},\qquad d_f' = 3d / 4, 
\end{equation}
\noindent 
which we support by extensive simulations and hyperscaling arguments (see Discussion). In light of the above, we advance a scaling theory for the fluctuations of the order parameters' mass close to criticality as follows. At short scales, $L<\xi'$, $\sigma(\mathcal{O})$ follows Eq.~\eqref{eq: fractal_dimension}, while at long scales, $L>\xi'$, the fluctuations are non-critical and satisfy the scaling law $\sigma(\mathcal{O}) \sim \sqrt{N} = L^{d/2}$ (see SI). Combining the above observation, we obtain the scaling function
\begin{equation}
    \frac{\sigma(\mathcal{O})}{L^{d/2} \xi'^{d/4}} = \mathcal{G}(L / \xi')
    \label{eq :scaling}
\end{equation}
where $\mathcal{G}(x)$ is constant for $x \gg 1$ and  $\mathcal{G}(x) \sim x^{d/4}$ for $x \ll 1$. Remarkably, we find that Eq.~\eqref{eq :scaling} is valid for both randomly interdependent percolation (Fig.~\ref{fig:nuprime}, Fig.~\ref{fig:dfprime}) and magnetization processes (Fig.~\ref{fig:spins}), both on spatial and on random networks (see SI), and it is satisfied for any dimension $d\geq2$, hinting at its universal nature. 

\underline{\textit{Interdependent percolation}}--To percolate our system of randomly interdependent lattices, we remove at random a fraction $1-p$ of the nodes belonging to one layer and let the cascading of failures to propagate back-and-forth between the layers. In finite systems, each realization is characterized by a distinct percolation threshold, $p_c$, and a critical mass of the mutual giant connected component (MGCC), $M_c = S_{\infty}^c L^d$, where $S_{\infty}^c$ is the relative fraction of nodes within the MGCC at criticality (Fig.~\ref{fig:Illustration}\textbf{b}). 
We find that their distributions fit a \textit{skewed Gaussian} (Fig.~\ref{fig:Illustration}\textbf{d,e}) with non-zero skewness~\cite{note2} $\gamma_1 = \langle x^3 \rangle$ and kurtosis $\kappa = \langle x^4 \rangle$ (see Fig.~\ref{fig:Illustration}\textbf{c},\textbf{d}). Here, $x = (y - \langle y \rangle) / \sigma(y)$ is the normalized parameter of the distribution and $y$ is the observable of interest ($p_c$ and $M_c$ in our case). 
The normal form of the two distributions indicate that the thermodynamic exponents $\beta',\,\gamma',\,\delta',\dots$ characterizing the critical fluctuations of the order parameter can be cast within the standard $\phi^4$-field theory~\cite{parisi1998statistical} and, as such, they belong to the mean-field Ising universality class (see Discussions). Exponents and scaling relations directly related to the dimensionality $d$ of the underlying system, such as $\nu'$ or $d_f'$, on the other hand, are more delicate to study since they can be strongly influenced by the existence of dangerous irrelevant variables altering the singular part of the system's free-energy density~\cite{fisher1983advanced}. To determine $\nu'$ and $d_f'$ of mixed-order percolation transitions in randomly interdependent lattices, we follow the method proposed in the above and analyze the finite-size scaling of $\sigma^2(p_c)$ and $\sigma^2(M_c)$. As shown in Fig.~\ref{fig:nuprime}\textbf{a}, the scaling of the width of the distribution $P(p_c)$ yields a correlation length exponent, $\nu'$, whose expression explicitly depends on the network's dimension and nicely agrees with the relation $\nu'=2/d$ (Fig.~\ref{fig:nuprime}\textbf{a} inset). To further corroborate the expression $\nu'=2/d$, we perform a data collapse of the distributions $P(p_c)$ obtained for different system's sizes which, in light of Eq.~\eqref{eq: xi_prime}, can be rescaled~\cite{bunde1991fractals} as 
\begin{equation}
    P(p_c) L^{-1/ \nu'} \sim \mathcal{F}\big[(p_c-\langle p_c \rangle)/L^{-1/ \nu'}\big], 
    \label{eq: pc_dist}
\end{equation}
where $\mathcal{F}(x)$ fits the profile of a skewed Gaussian. As shown in Fig.~\ref{fig:nuprime}\textbf{b} for $d = 2$, the data gathered over different system's sizes nicely collapse. 
\par
As anticipated in the above, the diverging correlation length $\xi'$ manifests physically the self-similar character of critical fluctuations on the order parameters' mass, $M_c$, at the hybrid percolation threshold. Indeed, as displayed in Fig.~\ref{fig:dfprime}\textbf{a}, the scaling advanced in Eq.~\eqref{eq: fractal_dimension} is nicely corroborated by means of extensive simulations and the fractal fluctuation dimension $d_f'=3d/4$ is observed over several decades in randomly interdependent lattices of dimensions ranging from $d=2$ up to $d=7$. The inset of Fig.~\ref{fig:dfprime}\textbf{a}, in particular, highlights the validity of the expression $d_f'=3d/4$ which we corroborate by performing a data collapse of the distribution $P(M_c)$ (see Fig.~\ref{fig:dfprime}\textbf{b} and details in caption) and further justify by means of hyperscaling arguments in the Discussion. 
An interesting implication of the above results is that one of the classical properties of continuous phase transitions, i.e.\ a ratio $\langle M_c \rangle / \sigma(M_c)$ independent on the system size \cite{bramwell2001magnetic}, breaks down in mixed-order transitions. In fact, since here the MGCC itself is now compact at criticality, $\langle M_c \rangle \sim L^d$ and the ratio scales therefore with the fluctuations co-dimension $\Delta' = d - d_f' = d/4$ as $\langle M_c \rangle / \sigma(M_c) \sim L^{\Delta'}$. 
\par
To complete the picture, we analyze the structure of fluctuations near the mixed-order percolation threshold. In light of Eq.~\eqref{eq :scaling}, we expect that when taking a small displacement $\Delta p=p-p_c$ from the abrupt threshold, the critical fluctuations of the MGCC's mass will be fractal with dimension $d_f'$ up to $\xi'$ and non-critical (i.e.\ purely Gaussian, see Fig.~S1 in the SI) otherwise. We support the above picture in Fig.~\ref{fig:dfprime}\textbf{c} with simulations in $d=2$ and $d=3$ (Fig.~\ref{fig:dfprime}\textbf{c}, inset) lattices. By rescaling the critical width $\Delta p$ via Eq.~\eqref{eq: xi_prime}, the crossover between the (critical) fractal fluctuations regime and the (non-critical) Gaussian regime is nicely seen. We further support the presence of such a critical crossover at the mixed-order percolation transition of interdependent random graphs (see Fig.~S4 in the SI).

\par
\underline{\textit{Interdependent magnetization}}-- To scrutinize the universality of the fractal fluctuations phenomenon at mixed-order phase transitions, we consider a model of interdependent Ising-spin networks where dependency couplings between layers are realized as thermal interactions~\cite{bonamassa2021interdependent}. 
We consider here the particular case of $d$-dimensional lattices modeled as in Fig.~\ref{fig:Illustration}\textbf{a}, where each node is endowed with an Ising spin $\sigma_i=\pm1$. Dependency couplings between the layers are modeled as local thermal feedback on the level of the flipping probability of spins (see Fig.~S2 in SI and discussions therein for details), which intertwine adaptively the stochastic (Monte-Carlo) dynamics of the two layers. 
In randomly interdependent spins networks, the average magnetization, $M = \sum_i \langle \sigma_i\rangle_\beta$ (where $\langle(\,\cdots)\rangle_\beta$ is a thermal average) undergoes a spontaneous mixed-order ferromagnetic-to-paramagnetic transition (Fig.~\ref{fig:spins}\textbf{a}) at a finite critical temperature $T_c\propto 1.25 d$ (see Fig.~\ref{fig:spins}\textbf{a}, inset) and for any dimension $d\geq2$, analogously to the abrupt collapse of the MGCC at $p_c$ in interdependent percolation. 
\par
We perform an analysis analogous to the one put forward in the above for interdependent percolation and determine the fluctuation correlation length exponent, $\nu'$, and its fractal dimension, $d_f'$, respectively from the scaling of $\sigma(T_c)$ and of $\sigma(M_c)$. Fig.~\ref{fig:spins}\textbf{b} shows the finite-size scaling of $\sigma(T_c)$, whose behavior nicely follows Eq.~\eqref{eq: nuprime} with $\nu' = 2/d$ (see Fig.~\ref{fig:spins}\textbf{b}, inset) and supports the existence of a diverging correlation length, Eq.~\eqref{eq: xi_prime}. Moving to the self-similar properties, we display in Fig.~\ref{fig:spins}\textbf{c} the scaling of $\sigma(M_c)$ and confirm the validity of the fractal fluctuation dimension $d_f' = 3d/4$ (Fig.~\ref{fig:spins}\textbf{c} inset) in agreement with Eq.~\eqref{eq: fractal_dimension}. Both our measurements of $\nu'$ and of $d_f'$ are further validated via their corresponding distributional collapse (see Fig.~S3\textbf{a},\textbf{b} in the SI for details). To close the picture, we show in Fig.~\ref{fig:spins}\textbf{d} the behavior of critical fluctuations near the mixed-order magnetization transition in $d=2$. We find also here excellent agreement with the scaling relation proposed in Eq.~\eqref{eq :scaling}, supporting further the universality of the crossover at hybrid transitions from the fractal ($L\ll\xi'$) to the non-fractal ($L\gg\xi'$) fluctuation regime. 

\par
\underline{\textit{Discussion}}-- In this Letter, we have unveiled the presence of self-similarity and of a diverging length scale, Eq.~\eqref{eq: xi_prime}, in the structure of critical fluctuations at mixed-order phase transitions. These critical phenomena are all the more surprising, as the order parameter at the hybrid transition is always compact and develops a discontinuity at the transition's threshold. We characterize these phenomena via the fluctuations' fractal dimension, $d_f'=3d/4$, and their corresponding correlation length exponent, $\nu'=2/d$, which we verify numerically for percolation and magnetization processes both {\em at}, Eqs.~\eqref{eq: nuprime},\,\eqref{eq: fractal_dimension}, and {\em close} to, Eq.~\eqref{eq :scaling}, the mixed-order transition's threshold. Since the critical fluctuations at the mixed-order transition are Gaussian-distributed (Fig.~\ref{fig:nuprime}\textbf{b}, Fig.~\ref{fig:dfprime}\textbf{b} and Fig.~S3\textbf{a},\,\textbf{b} in the SI), a hyperscaling hypothesis can be put forward so to interpret the scaling exponents $(\nu',d_f')$ in terms of the thermodynamic ones $\beta'=1/2$, $\gamma'=1$, etc.\ featuring the mean-field Ising universality class~\cite{lee2016hybrid}. Indeed, one can readily verify that the hyperscaling relations $d\nu'=2\beta'+\gamma'$ and $d_f'=d-\beta'/\nu'$ are identically satisfied. Since our results hold a universal character, we expect the phenomenon of fractal fluctuations to be experimentally observable in other models undergoing mixed-order phase transitions, for example in colloidal crystals~\cite{alert2017mixed}, networks of active gels~\cite{sheinman2015anomalous}, multilayer superconductors~\cite{bonamassa2022superconductors}, as well as in physical and biological systems belonging to the universality class of random first-order phase transitions~\cite{kirkpatrick2015colloquium}. The latter, in particular, hints at an intriguing cross-disciplinary connection which deserves further research. 


\par
\underline{\textit{Acknowledgements}}-- We thank the Israel Science Foundation, the NSF-BSF Grant No.\ 2019740, the EU H2020 project RISE (Project No. 821115) and the EU H2020 DIT4TRAM for financial support.

\bibliographystyle{unsrt}
\bibliography{mybib}

\begin{thebibliography}{10}
\bibitem{note2}
Interestingly, the skeweness in the distribution indicates that the critical
  fluctuations are biased toward the low stable state with larger deviations
  from the mean compared to the metastable one.
\end{thebibliography}

\newpage
\begin{figure}\vspace*{-0.5cm}
\centering 
{\hspace*{-0.75cm} \includegraphics[scale = 0.9, page = 1]{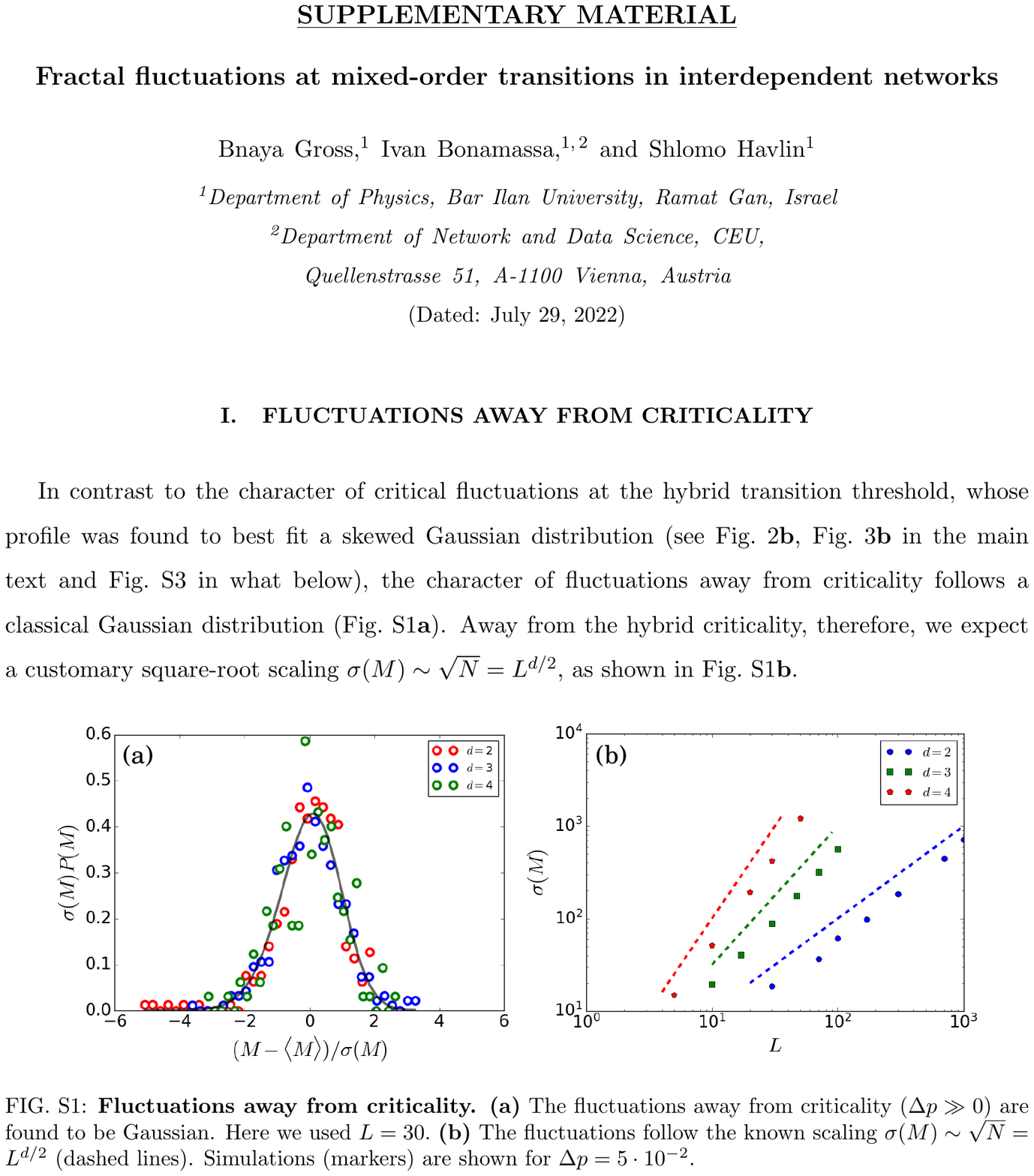}}
\end{figure}

\newpage
\pagenumbering{gobble}
\begin{figure}\vspace*{-0.5cm}
\centering
{\hspace*{-0.75cm}\includegraphics[scale = 0.9, page = 2]{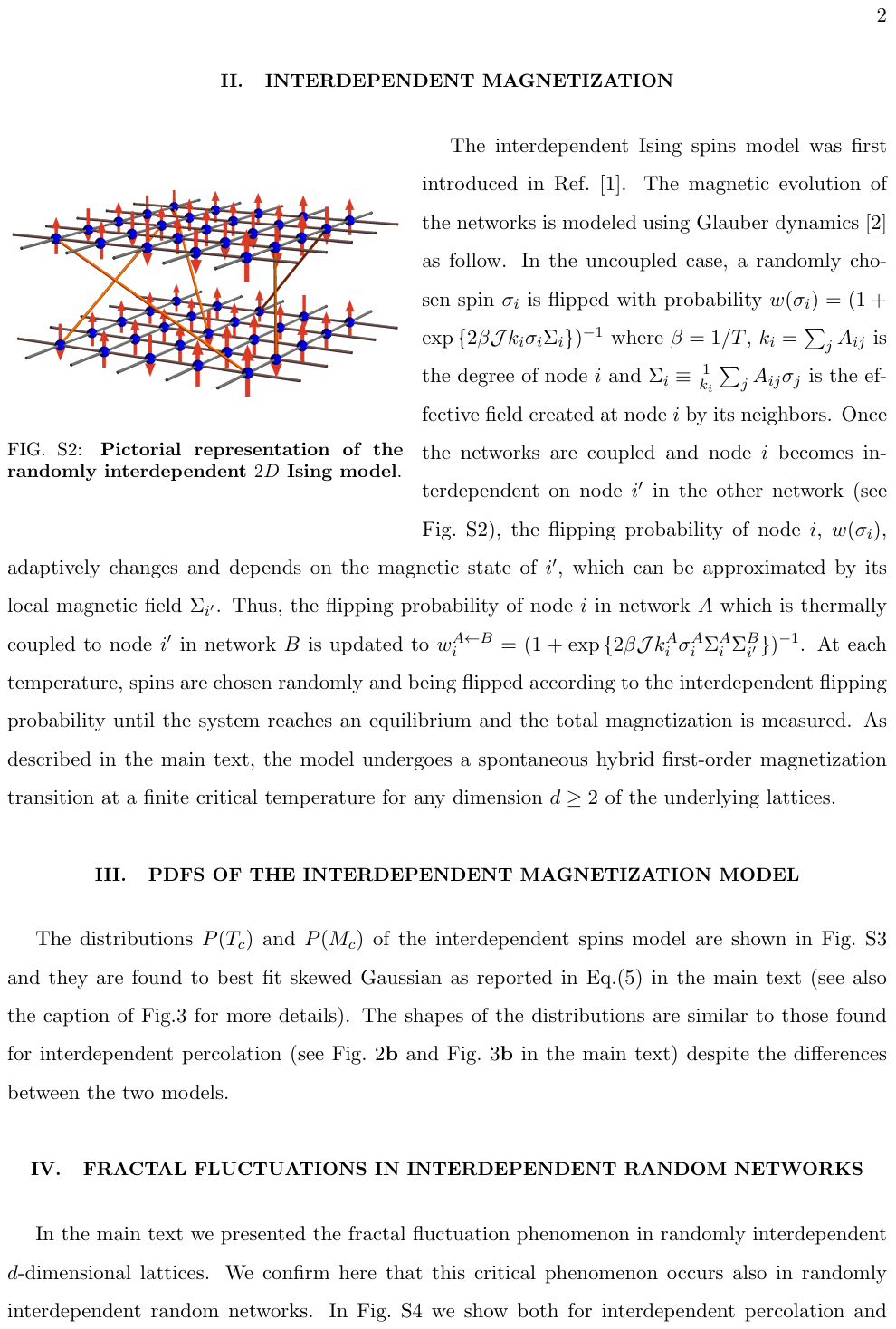}}
\end{figure}

\newpage
\pagenumbering{gobble}
\begin{figure}\vspace*{-0.5cm}
\centering
{\hspace*{-0.75cm}\includegraphics[scale = 0.9, page = 3]{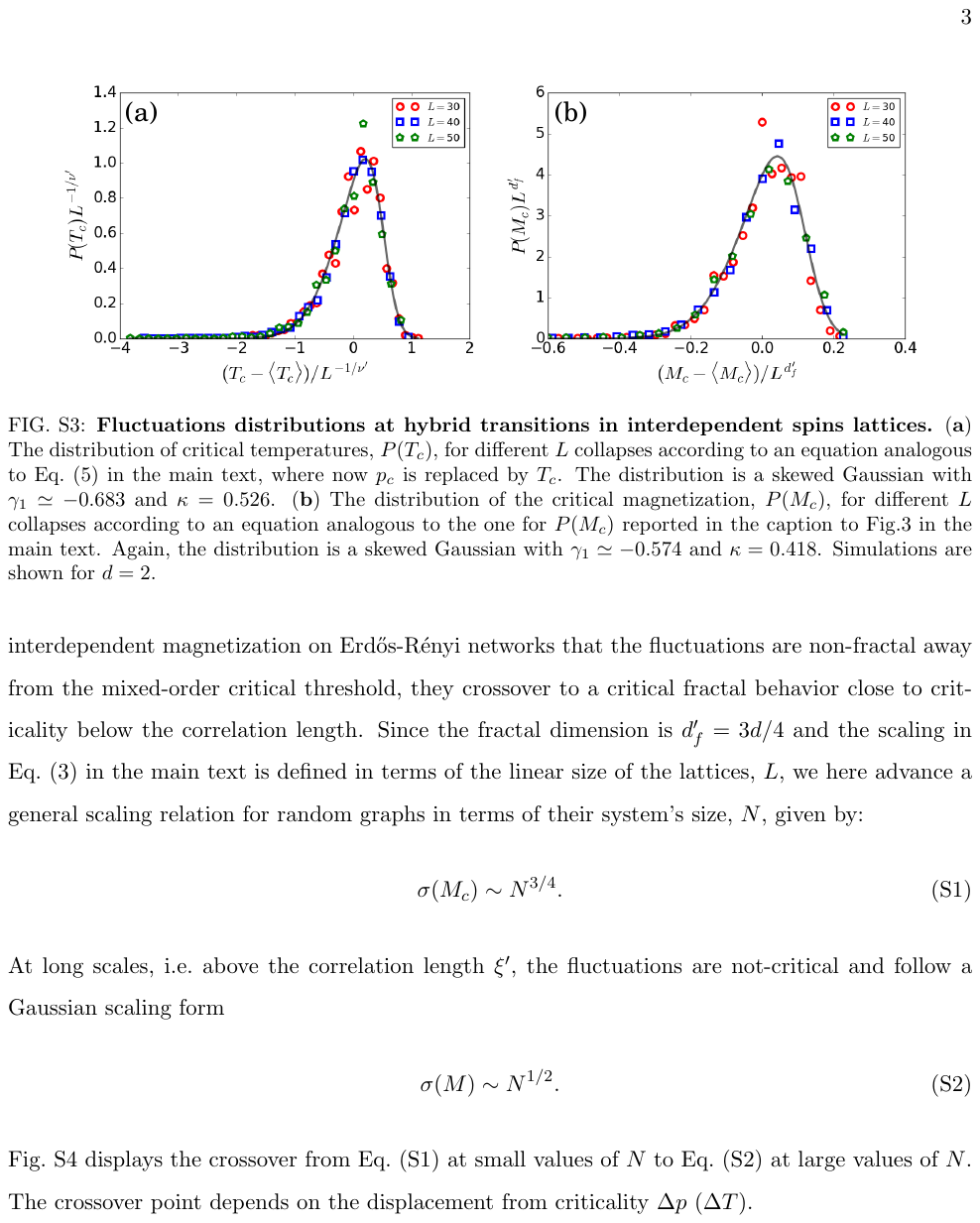}}
\end{figure}

\newpage
\pagenumbering{gobble}
\begin{figure}\vspace*{-0.5cm}
\centering
{\hspace*{-0.75cm}\includegraphics[scale = 0.9, page = 4]{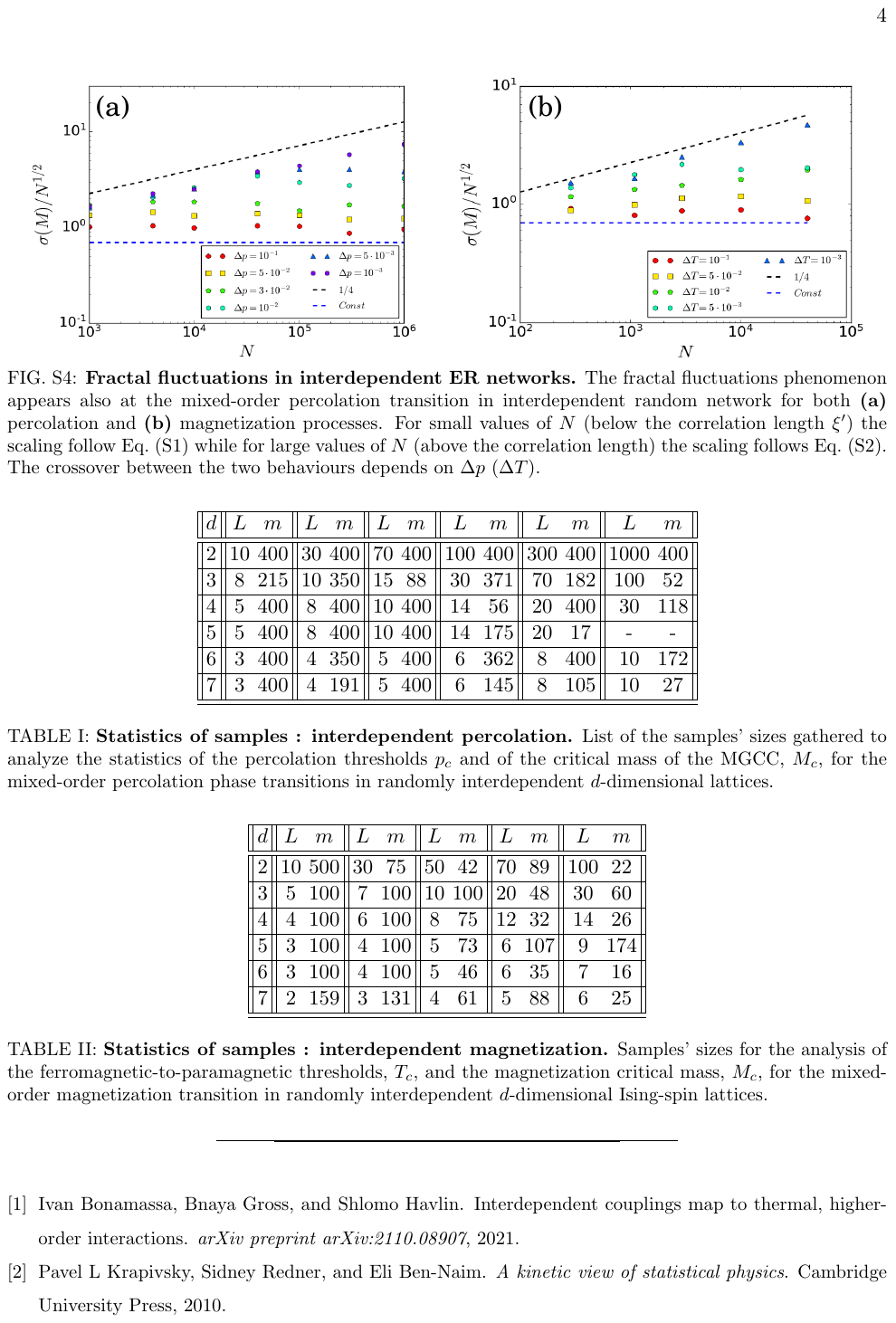}}
\end{figure}
\end{document}